\documentstyle[seceq,preprint,graphicx,dcolumn,mbf,epsf]{ptptex}

\title{
Chiral Sigma Model with Pion Mean Field\\ in Finite Nuclei }

\author{
Yoko Ogawa$^{1}$\footnote{E-mail address:
ogaway@rcnp.osaka-u.ac.jp}, Hiroshi Toki$^{1,2,}$\footnote{E-mail address:
toki@rcnp.osaka-u.ac.jp}, Setsuo Tamenaga$^{1}$, Hong Shen$^{3}$,
\\ Atsushi Hosaka$^{1}$, Satoru Sugimoto$^{2}$
, and Kiyomi Ikeda$^{2}$
}

\inst{
$^1$Research Center for Nuclear Physics (RCNP), Osaka University,\\
Ibaraki, Osaka 567-0047, Japan
\\
$^2$Institute of Chemical and Physical Research (RIKEN),\\
Wako, Saitama 351-0198, Japan\\
$^3$Department of physics, Nankai University, Tianjin 300071, China }

\recdate{
}

\abst{
The properties of infinite matter and finite nuclei are studied by
using the chiral sigma model in the framework of the relativistic
mean field theory. We reconstruct an extended chiral sigma model
in which the omega meson mass is generated dynamically by the
sigma condensation in the vacuum in the same way as the nucleon
mass. All the parameters of chiral sigma model are essentially
fixed from the hadron properties in the free space. In nuclear
matter, the saturation property comes out right, but the
incompressibility is too large and the scalar and vector
potentials are about a half of the phenomenological ones,
respectively.  This fact is reflected to the properties of finite
nuclei.  We calculate N = Z even-even mass nuclei between N = 16
and N = 34. The extended chiral sigma model without the pion mean
field leads to the result that the magic number appears at N = 18
instead of N = 20 and the magic number does not appear at N = 28
due to the above mentioned nuclear matter properties. The latter
problem, however, could be removed by the introduction of the
finite pion mean field with the appearance of the magic number at
N = 28. We find that the energy differences between the spin-orbit
partners are reproduced by the finite pion mean field which is
completely a different mechanism from the standard spin-orbit
interaction.}

\begin{document}

\maketitle

\section{Introduction}

Chiral symmetry is known to be the most important symmetry in
the hadron physics.  This is because the quantum chromo-dynamics
(QCD) is the underlying theory of the strong interaction, in which
the up and the down quarks have essentially zero masses. Chiral symmetry
governs the quark dynamics.  In the real world, the quarks are
confined and chiral symmetry is spontaneously broken.  As the
Nambu-Goldstone boson of the spontaneous breaking of chiral symmetry,
the pion emerges with almost zero mass.

At the hadron level, chiral symmetry was described nicely in the
linear sigma model introduced by Gell-Mann and Levy
\cite{gellmann}. Its non-linear version was proposed by
Weinberg\cite{weinberg}. Chiral symmetry and the generation of the
hadron mass were described clearly in the Nambu-Jona-Lasinio
Lagrangian with fermion fields \cite{nambu}. These Lagrangians
have been used for various phenomena in hadron physics. We find
good description of the pion-nucleon properties in terms of these
Lagrangians \cite{lee}.  The pion, which was introduced by Yukawa
as the mediator of the nucleon-nucleon interaction, received its
foundation through the spontaneous chiral symmetry breaking
\cite{yukawa35}.

It is then very natural to use the chiral sigma model Lagrangian
for the description of nuclei. This was performed by several
groups in the relativistic mean field
approximation\cite{walecka74,boguta1,boguta2,savushkin1,savushkin2}.
It was found that the use of chiral sigma model in its original
form was not satisfactory for the description of nuclear matter.
An interesting way out was proposed by Boguta et al., who
introduced a dynamical generation of the omega meson mass in the
same way as the nucleon mass\cite{boguta1}. They were able to
reproduce the saturation properties of infinite matter with the
extended chiral sigma (ECS) model.  However, the effective mass
came out to be large and the incompressibility to be very large.
The extended chiral sigma model was applied to finite nuclei by
Savushkin et al.\cite{savushkin1,savushkin2}.  The binding
energies came out to be reasonable, but the spin-orbit splitting
was too small in the RMF framework.

Recently, an interesting proposal was made on the role of pion in
finite nuclei by some of the present authors\cite{toki}.  They
made the relativistic mean field calculations with finite pion
mean field using the TM1 parameter set \cite{sugahara94}.  In
order to treat the pion mean field, they developed the formalism,
in which parity mixed single particle states were introduced. With
the use of the free space pion-nucleon coupling constant, they
found that the pion mean field becomes finite, especially the
effect appears favorably for the jj-closed shell nuclei, and the
mass dependence of the energy gain associated with the pion
behaves as the nuclear surface, $\langle V_\pi \rangle \sim
A^{2/3}$. Hence, the name, surface pion condensation, was
introduced for this phenomenon.

In this paper, we would like to study the properties of infinite
matter in terms of the ECS model by analyzing the non-linear
equation of motion for the sigma field and obtain the saturation
property of nuclear matter. We apply the ECS model to finite
nuclei and study the properties of the binding energies and the
single particle properties.  Since the role of the finite pion
mean field on the binding energies and the spin-orbit splitting
has been demonstrated in the recent publication\cite{toki}, we
take the formalism to treat the finite pion mean field in the RMF
framework in the ECS model for the calculation of finite nuclei.
We would like to study the appearance of the spin-orbit splitting
due to the pion mean field by studying carefully the single
particle spectra of finite nuclei.

In section~\ref{sec: rmf}, we discuss the RMF formalism with the
pion mean field.  In section \ref{sec: numinf}, we study the
saturation property of infinite nuclear matter with the original
chiral sigma model and with the extended chiral sigma model.  In
section \ref{sec: numfin}, we study finite nuclei with the
extended chiral sigma model without introducing yet the pion mean
field and further study the properties of single particle states.
We introduce then in section 5 the finite pion mean field and
discuss the mechanism of the appearance of the magic number effect
and the energy splittings between the spin-orbit partners. We
summarize the present study in section~\ref{sec: conc} together
with the statements for the further study.

\section{Chiral sigma model in the relativistic mean field theory}
\label{sec: rmf}

We start with the linear sigma model with the omega meson field,
which is defined by the following Lagrangian\cite{gellmann},
\begin{eqnarray}
{\cal L}_{\sigma\omega} & = &
\bar{\Psi}(i\gamma_{\mu}\partial^{\mu}
                          - g_{\sigma}(\sigma + i\gamma_{5}{\vec\tau}\cdot{\vec\pi})
                          - g_{\omega}\gamma_{\mu}\omega^{\mu})\Psi \\ \nonumber
& + &  \frac{1}{2} \partial_{\mu}\sigma \partial^{\mu}\sigma
                 + \frac{1}{2} \partial_{\mu}{\vec\pi} \partial^{\mu}{\vec\pi}
                 - \frac{\mu^{2}}{2}(\sigma^2 + {\vec\pi}^2)
                 - \frac{\lambda}{4}(\sigma^2 + {\vec\pi}^2)^2 \\ \nonumber
& - &  \frac{1}{4}\omega_{\mu\nu}\omega^{\mu\nu}
                 + \frac{1}{2}{\widetilde{g_{\omega}}}^2
                   (\sigma^2 + {\vec\pi}^2)\omega_{\mu}\omega^{\mu} \\ \nonumber
\hspace{3cm} & + & \varepsilon\sigma.
\end{eqnarray}
The fields $\Psi$, $\sigma$ and $\pi$ are the nucleon, sigma and
the pion fields.  $\mu$ and $\lambda$ are the sigma model coupling
constants.  Here we have introduced the explicit chiral symmetry
breaking term, $\varepsilon\sigma$, and in addition the mass
generation term for the omega meson due to the sigma meson
condensation as the case of the nucleon mass in the free space
\cite{boguta1}.  The $\sigma-\omega$ coupling term of this structure
may be derived from the bosonization\cite{hosaka} of the Nambu-Jona-Lasinio model
\cite{nambu}.


In a finite nuclear system, it is believed to be essential to use
the non-linear representation of the chiral symmetry.  This is
because the pseudoscalar pion-nucleon coupling in the linear sigma
model makes the coupling of positive and the negative energy
states extremely strong and we have to treat the negative energy
states very carefully. We can derive the non-linear sigma model by
introducing new variables and making a suitable transformation,
\begin{eqnarray}
\sigma + i{\vec\tau}\cdot{\vec\pi} & = & \rho{\rm U}, \hspace{1cm}
{\rm U} = e^{i{\vec\tau}\cdot{\vec\pi} / f_{\pi}} \\ \nonumber \sigma +
i\gamma_{5}{\vec\tau}\cdot{\vec\pi} & = & \rho{\rm U}_{5},
\hspace{1cm} {\rm U}_{5} =
e^{i\gamma_{5}{\vec\tau}\cdot{\vec\pi} / f_{\pi}},
\end{eqnarray}
We further implement the Weinberg transformation for the nucleon field
as $\psi = \sqrt{{\rm U}_{5}}\Psi$.  We obtain then the
sigma-omega model Lagrangian in non-linear representation,
\begin{eqnarray}
{\cal L}'_{\sigma\omega} & = &
 \bar\psi( i\gamma_{\mu}\partial^{\mu}
             - g_{\sigma}\rho
             - \gamma_{\mu}v^{\mu}
             - \gamma_{5}\gamma_{\mu}a^{\mu}
             - g_{\omega}\gamma_{\mu}\omega^{\mu} )\psi \\ \nonumber
& + &     \frac{1}{2}\partial_{\mu}\rho\partial^{\mu}\rho
        + \frac{\rho^2}{4}{\rm tr}\partial_{\mu}{\rm U}
                                  \partial^{\mu}{\rm U}^{\dagger}
        - \frac{\mu^2}{2}\rho^2
        - \frac{\lambda}{4}\rho^4 \\ \nonumber
& - &     \frac{1}{4}\omega_{\mu\nu}\omega^{\mu\nu}
        + \frac{1}{2}{\widetilde{g_{\omega}}}^2\rho^2
                      \omega_{\mu}\omega^{\mu}  +  \varepsilon\rho\frac{1}{2}( {\rm U}
                      + {\rm U}^{\dagger}).
\end{eqnarray}
In the above Lagrangian the vector field, $v^{\mu}$, and the axial
vector field, $a^{\mu}$, contain the pion terms. The vector and
the axial vector fields are expanded in terms of the pion field as,
\begin{eqnarray}
v^{\mu} & = & \frac{-i}{8{f_{\pi}}^2}
              (  {\vec\tau}\cdot{\vec\pi}
                 {\vec\tau}\cdot\partial^{\mu}{\vec\pi}
               - {\vec\tau}\cdot\partial^{\mu}{\vec\pi}
                 {\vec\tau}\cdot{\vec\pi} )
           + \cdot\cdot\cdot, \\ \nonumber
a^{\mu} & = & \frac{1}{2f_{\pi}}
              {\vec\tau}\cdot\partial^{\mu}{\vec\pi}
                          + \cdot\cdot\cdot. \nonumber
\end{eqnarray}
The kinetic term is expanded as follows,
\begin{equation}
\frac{\rho^2}{4}{\rm tr}\partial_{\mu}{\rm U}
                        \partial^{\mu}{\rm U}^{\dagger} =
\frac{\rho^2}{2{f_{\pi}}^2}\partial_{\mu}{\vec\pi}
                           \partial^{\mu}{\vec\pi}
+ {\cal O}({\vec\pi}^4) + {\cal O}({\vec\pi}^6) + \cdot\cdot\cdot,
\nonumber
\end{equation}
and the explicitly chiral symmetry breaking term is expanded as
follows,
\begin{equation}
\varepsilon\rho\frac{1}{2}({\rm U} + {\rm U}^{\dagger}) =
\varepsilon\rho( 1 - \frac{1}{2{f_{\pi}}^2}{\vec\pi}^2
                    + \frac{1}{24{f_{\pi}}^4}{\vec\pi}^4
                    + \cdot\cdot\cdot ). \nonumber
\end{equation}
We take now the lowest order terms in the pion filed and truncate
higher order terms.  The resulting Lagrangian is written as,
\begin{eqnarray}
{\cal L}'_{\sigma\omega} & = & \bar{\psi}(
i\gamma_{\mu}\partial^{\mu}
            - g_{\sigma}\rho
            - \frac{1}{2f_{\pi}}
              \gamma_{5}\gamma_{\mu}
             {\vec\tau}\cdot\partial^{\mu}{\vec\pi}
            - g_{\omega}\gamma_{\mu}\omega^{\mu} )\psi \\ \nonumber
& + &    \frac{1}{2}\partial_{\mu}\rho\partial^{\mu}\rho
       + \frac{1}{2}\frac{\rho^2}{{f_{\pi}}^2}
                    \partial_{\mu}{\vec\pi}\partial^{\mu}{\vec\pi}
       - \frac{\mu^2}{2}\rho^2
       - \frac{\lambda}{4}\rho^4 \\ \nonumber
& - &    \frac{1}{4}\omega_{\mu\nu}\omega^{\mu\nu}
       + \frac{1}{2}{\widetilde{g_{\omega}}}^2\rho^2
                     \omega_{\mu}\omega^{\mu}  +   \varepsilon\rho( 1 -
\frac{1}{2{f_{\pi}}^2}{\vec\pi}^2 ).
\end{eqnarray}

We now take the vacuum expectation value for the $\rho$ field as
$f_{\pi}$, which is determined by the pion decay rate\cite{lee},
\begin{equation}
\langle \rho \rangle_{0} = f_{\pi}.
\end{equation}
A new fluctuation field $\varphi$ may be defined by the equation,
\begin{equation}
\rho = f_{\pi} + \varphi.
\end{equation}
We shall now rewrite the Lagrangian (2.7) in terms of the new field
$\varphi$,
\begin{eqnarray}
{\cal L}'_{\sigma\omega} & = &
\bar{\psi}(i\gamma_{\mu}\partial^{\mu}
            - g_{\sigma}f_{\pi}
            - g_{\sigma}\varphi
            - \frac{1}{2f_{\pi}}
              \gamma_{5}\gamma_{\mu}
             {\vec\tau}\cdot\partial^{\mu}{\vec\pi}
            - g_{\omega}\gamma_{\mu}\omega^{\mu} ){\psi} \\ \nonumber
& + &   \frac{1}{2}\partial_{\mu}\varphi\partial^{\mu}\varphi
      + \frac{1}{2}( 1 + \frac{\varphi}{f_{\pi}} )^2
                   \partial_{\mu}{\vec\pi}\partial^{\mu}{\vec\pi}
      - \frac{\mu^2}{2}( f_{\pi} + \varphi )^2
      - \frac{\lambda}{4}( f_{\pi} + \varphi )^4 \\ \nonumber
& - &   \frac{1}{4}\omega_{\mu\nu}\omega^{\mu\nu}
      + \frac{1}{2}{\widetilde{g_{\omega}}}^2( f_{\pi}+ \varphi )^2
                                  \omega_{\mu}\omega^{\mu} \\ \nonumber
& + &   \varepsilon( f_{\pi} + \varphi )
                   ( 1 - \frac{1}{2{f_{\pi}}^2}{\vec\pi}^2 ).
\end{eqnarray}
Here, we have dropped a non-essential c-number constant in the above
expression. We find the term ``$\varphi / f_{\pi}$'' is small and
drop it as follows,

\begin{eqnarray}
\frac{1}{2}( 1 + \frac{\varphi}{f_{\pi}} )^2
\partial_{\mu}{\vec\pi}\partial^{\mu}{\vec\pi} & \approx &
\frac{1}{2}\partial_{\mu}{\vec\pi}\partial^{\mu}{\vec\pi}, \\
\nonumber ( 1 + \frac{\varphi}{f_{\pi}} )
\frac{1}{2}\frac{\varepsilon}{f_{\pi}}{\vec\pi}^2 & \approx &
\frac{1}{2}\frac{\varepsilon}{f_{\pi}}{\vec\pi}^2. \nonumber
\end{eqnarray}
We have to make the dangerous term, the term linear in $\varphi$, zero,
which leads to the energy minimum condition.
\begin{eqnarray}
(\varepsilon - {m_{\pi}}^2f_{\pi} )\varphi & \longrightarrow & 0,
\\ \nonumber {m_{\pi}}^2 & = & \frac{\varepsilon}{f_{\pi}}.
\end{eqnarray}
Finally the Lagrangian for the new field $\varphi$ within the
above approximations is given as follows,
\begin{eqnarray}
{\cal L}'_{\sigma\omega} & = & \bar{\psi}(
i\gamma_{\mu}\partial^{\mu}
            - M
            - g_{\sigma}\varphi
            - \frac{1}{2f_{\pi}}
              \gamma_{5}\gamma_{\mu}
             {\vec\tau}\cdot\partial^{\mu}{\vec\pi}
            - g_{\omega}\gamma_{\mu}\omega^{\mu} ){\psi} \\ \nonumber
& + &   \frac{1}{2}\partial_{\mu}\varphi\partial^{\mu}\varphi
      - \frac{1}{2}{m_{\sigma}}^2\varphi^2
      - \lambda{f_{\pi}}\varphi^3
      - \frac{\lambda}{4}\varphi^4 \\ \nonumber
& + &   \frac{1}{2}\partial_{\mu}{\vec\pi}\partial^{\mu}{\vec\pi}
      - \frac{1}{2}{m_{\pi}}^2{\vec\pi}^2 \\ \nonumber
& - &   \frac{1}{4}\omega_{\mu\nu}\omega^{\mu\nu}
      + \frac{1}{2}{m_{\omega}}^2\omega_{\mu}\omega^{\mu}
      + {\widetilde{g_{\omega}}}^2f_{\pi}\varphi
                       \omega_{\mu}\omega^{\mu}
      + \frac{1}{2}{\widetilde{g_{\omega}}}^2\varphi^2
                                  \omega_{\mu}\omega^{\mu},
\end{eqnarray}
where we set $M = g_{\sigma}f_{\pi}$, ${m_{\pi}}^2 = \mu^2 +
\lambda{f_{\pi}}^2$, ${m_{\sigma}}^2 = \mu^2 +
3\lambda{f_{\pi}}^2$ and $m_{\omega}  =
\widetilde{g_{\omega}}f_{\pi}$.  The effective mass of the nucleon
and omega meson are given by $M^{\ast} = M + g_{\sigma}\varphi$ and
${m_{\omega}}^{\ast} = m_{\omega} +
\widetilde{g_{\omega}}\varphi$, respectively. We take the
following masses and the pion decay constant as,
 $M = 939 \ \rm{MeV}$,
 $m_{\omega} = 783 \ {\rm MeV}$,
 $m_{\pi} = 139 \ {\rm MeV}$, and
 $f_{\pi} = 93 \ {\rm MeV}$.
Then, the other parameters can be fixed automatically by the
following relations, $g_{\sigma}  = M/f_\pi  = 10.1$ and
$\widetilde{g_{\omega}} = m_{\omega}/f_{\pi} = 8.42$. The strength
of the cubic and quadratic sigma meson self-interactions depends
on the sigma meson mass through the following relation, $\lambda =
({m_\sigma}^2 - {m_\pi}^2)/2{f_\pi}^2$, in the chiral sigma model.
The mass of the sigma meson, $m_\sigma$, and the coupling constant
of omega and nucleon, $g_\omega$, are the free parameters. If we
use the KSFR relation for the omega meson\cite{kawa,riazuddin},
and the additional relation from the Nambu-Jona-Lasinio model, the
mass of the omega meson is related to the pion decay constant by
$m_\omega = (2\sqrt{2}/3) f_\pi g_\omega$.  The factor
$(2\sqrt{2}/3)$ stems from the $g_\omega = (3/2) g$, where $g$ is
the universal coupling constant for the vector
meson\cite{hosaka1,kaymakcalan}. As we see below, this KSFR
relation is very well satisfied in the present model within 6 \%.

\section{Extended chiral sigma model for infinite matter}\label{sec: numinf}

We apply first the extended chiral sigma model to infinite matter.
It is important to reproduce the saturation properties of infinite
nuclear matter first.  Otherwise, we do not get convergence due to
the multiple solutions in the Hartree calculation for finite
systems. We assume that the pion mean field vanishes in infinite
matter.  Hereafter we write the scalar meson field $\varphi$ in the
Lagrangian (2.13) as $\sigma$, since $\sigma$ is used usually as
the scalar meson field in the relativistic mean field theory. The equations
of motion for the nucleon field and the meson fields are written as,
\begin{eqnarray}
( i\gamma_{\mu}\partial^{\mu} - M - g_{\sigma}\sigma -
g_{\omega}\gamma^{0}\omega )\psi & = & 0, \\ \nonumber
{m_{\sigma}}^{2}\sigma + 3\lambda f_{\pi}\sigma^2 +
\lambda\sigma^3 - {\widetilde{g_{\omega}}}^{2}f_{\pi}\omega^2 -
{\widetilde{g_{\omega}}}^{2}\sigma\omega^2 & = &
-g_{\sigma}\rho_{s}, \\ \nonumber
{m_{\omega}}^{2}\omega +
2{\widetilde{g_{\omega}}}^{2}f_{\pi}\sigma\omega +
{\widetilde{g_{\omega}}}^{2}\sigma^{2}\omega & = &
g_{\omega}\rho_{v}.
\end{eqnarray}
with
\begin{eqnarray}
\rho_{s} & = & \frac{4}{(2\pi)^3}\int^{k_{\rm F}} d^{3}k
\frac{M^{\ast}}{\sqrt{k^{2} + {M^{\ast}}^{2}}} \\ \nonumber & = &
\frac{M^{\ast}}{\pi^2} {\Bigl \{ } k_{\rm F}\sqrt{{k_{\rm F}}^2 +
{M^{\ast}}^2} - {M^{\ast}}^{2} \log{\Bigl|}\frac{k_{\rm F} +
\sqrt{ {k_{\rm F}}^2 + {M^{\ast}}^2 }}{M^{\ast}}{\Bigr|} {\Bigr \}
},
\end{eqnarray}
$\rho_{v}  = 2{k_{\rm F}}^3/(3{\pi}^2)$ and the effective mass of the
nucleon $M^{\ast} = M + g_{\sigma}\sigma$.  We note here that now
the equations of motion of the sigma and omega mesons are
coupled due to the dynamical mass generation term of the omega
meson.  This sigma-omega coupling plays an important role to
obtain reasonable equation of state of nuclear matter.

We discuss first the original chiral sigma model for the nuclear
matter calculation \cite{boguta1}.  In this case, there is no
coupling between the equations for the sigma and the omega fields.
The equation for $\sigma$ is a third order algebraic equation of
the sigma together with minus of the scalar coupling times the
scalar-density, $-g_\sigma \rho_{s}$, which is a function of the
sigma field for a fixed density,
\begin{eqnarray}
m_\sigma^2 \sigma + 3\lambda f_{\pi} \sigma^2 + \lambda \sigma^3 &
=  & -g_\sigma \rho_{s}.
\end{eqnarray}
The right hand side increases with decreasing the sigma field and
changes sign near $\sigma = - f_{\pi} \sim -0.5$ fm$^{-1}$. We
shall focus on the solution above the crossing point, until where
the effective mass of the nucleon is positive.  Below a certain
density there appears only one solution, while above this density
there appear three solutions. We get multiple solutions as
discussed above. For each solution, there is a corresponding
energy, which is not a smooth function of the density. Hence, we
are not able to get a good behavior for the equation of state with
the original chiral sigma model.

The way out to get a good nuclear matter property was suggested by
Boguta, who introduced the dynamical omega meson
term\cite{boguta1}. The omega mass appears due to the dynamical
chiral symmetry breaking and hence there is a coupling between the
sigma and the omega fields. We use this extended chiral sigma
model for nuclear matter. The additional term provides a pole at
the effective nucleon mass being zero, $\sigma = - f_{\pi}$, as
shown in Fig. 1.  Due to this reason we find a solution at a small
sigma value for each density continuously from zero. We are
therefore able to obtain a reasonable energy per particle in the
entire density region for infinite matter.

\begin{figure}[h]
\epsfxsize = 10 cm \centerline{\epsfbox{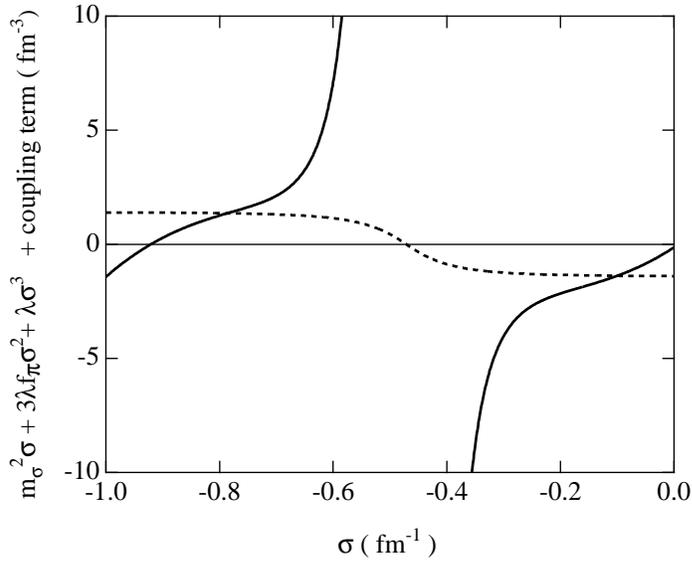}}
\caption[Fig.1]{The equation for $\sigma$ with the $\sigma-\omega$
coupling term for the case of $\rho=0.141$ fm$^{-3}$ in the
extended chiral sigma model.  There is one solution for each
density continuously from the zero density.}
\end{figure}

In Fig. 2 we provide the energy per particle of nuclear matter as
a function of the density for the extended chiral sigma model. We
take the parameters of the chiral sigma model from the properties
of mesons as pion mass, $m_\pi$, omega meson mass, $m_\omega$,
pion decay constant, $f_\pi$.  The free parameters, $m_\sigma$ and
$g_\omega$, are adjusted to provide the saturation property in the
case of the extended chiral sigma model.  We have fixed the free
parameters as, $m_{\sigma}$ = 777 MeV, and $g_{\omega}$ = 7.03.
Then, the strength of the cubic and quadratic sigma meson
self-interaction are fixed as $\lambda$ = 33.8. The saturation
properties are the density, $\rho$ = 0.141 fm$^{-3}$, and the
energy per particle, $E/A$ = -16.1 MeV.  We find in this case the
incompressibility, $K$ = 650 MeV. The sigma meson mass chosen here
is larger than that used in one boson exchange potential, which is
around 500 MeV. If we take 500 MeV as the sigma meson mass, the
attractive force becomes strong and the saturation curve becomes
deep. We adjust then the omega-nucleon coupling constant,
$g_\omega$, to reproduce the binding energy per particle.  The
energy minimum point appears at quite a small density, $\rho$ =
0.053 fm$^{-3}$. The saturation condition is not satisfied
simultaneously both for the density and binding energy per
particle using this meson mass. It is interesting to note that the
value $m_\sigma = 777$ MeV is very close to the one when the
chiral mixing angle is chosen at 45$^{\circ}$ in the generalized
chiral model; $m_\sigma \approx m_\rho$\cite{weinberg}.

\begin{figure}[h]
\epsfxsize = 10 cm \centerline{\epsfbox{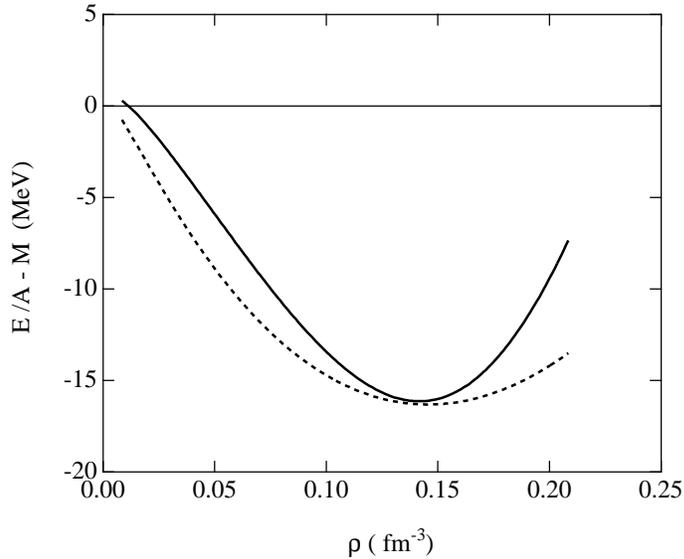}}
\caption[Fig.2]{The energy per particle of infinite nuclear matter
as a function of the density for the extended chiral sigma model
(solid curve). As a reference the energy per particle in the RMF
theory with the TM1 parameter set, RMF(TM1), is provided by dashed
curve.}
\end{figure}

\begin{figure}[h]
\epsfxsize = 10 cm \centerline{\epsfbox{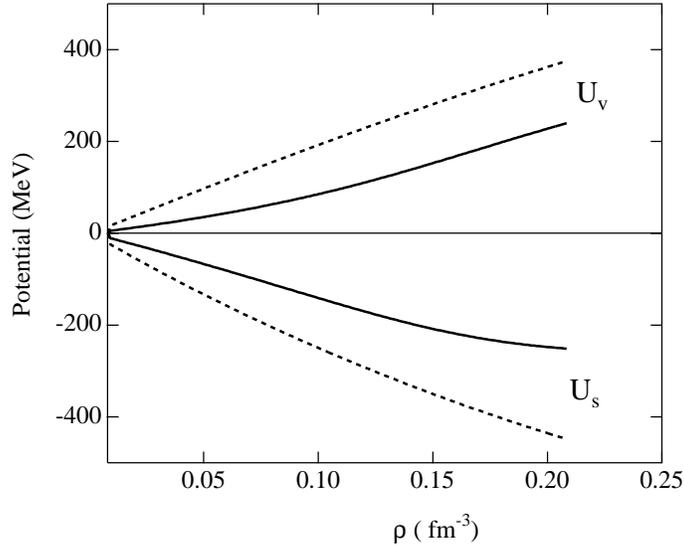}}
\caption[Fig.3]{The scalar and vector potentials are plotted as a
function of the density for the extended chiral sigma model shown
by solid curve and those for RMF(TM1) by dashed curve.}
\end{figure}

As a comparison, the energy per particle of the mean field result
with the TM1 parameter set is shown together with the present
result \cite{sugahara94}. The RMF(TM1) calculation reproduces the
results of the relativistic Brueckner-Hartree-Fock calculation
\cite{brockmann90}.  We see that the present equation of state is
much harder than the one of RMF(TM1). The incompressibility comes
out to be 650 MeV, while it is 281 MeV for TM1. In Fig. 3 we plot
the vector and the scalar potentials and compare with the ones of
RMF(TM1). The values are about a half of the case of the TM1
parameter set. This is because the extended chiral sigma model has
solutions at smaller sigma values than those for RMF(TM1).

We would like to note the consequence of the smaller absolute
values of the scalar and the vector potentials in finite nuclei as
shown in Fig. 3. The summation of the absolute values of the
scalar and the vector potentials is directly related with the
spin-orbit potential of finite nuclei.  Hence, the fact that these
absolute values are about a half of the values of RMF(TM1)
indicates that the spin-orbit splitting for finite nuclei will
come out to be about a half of the necessary spin-orbit
splittings.

\section{Extended chiral sigma model for finite nuclei}\label{sec: numfin}

We are now in the position to apply the extended chiral sigma
(ECS) model, which is able to provide the saturation property with
the above mentioned features, to finite nuclei. For this purpose
we take the N = Z even-even mass nuclei to avoid the complication
coming from the isovector part of the nucleon-nucleon interaction.
We calculate these nuclei using the RMF framework with the ECS
Lagrangian and compare the results with those of the standard RMF
calculation with the TM1 parameter set.  Since the role of the
pion mean field on the binding energy and the spin-orbit
interaction has been demonstrated in Ref. \cite{toki} for finite
nuclei, we shall introduce the RMF formalism on the treatment of
the finite pion mean field and study the effect of the finite pion
mean field on the nuclear properties.

We write here the RMF equations for the finite nuclei with the
pion mean field.  The Euler-Lagrange equation gives us the Dirac
equation for the nucleon:
\begin{equation}
(-i{\vec\alpha}\cdot\nabla
      + \gamma_{0}( M + g_{\sigma}\sigma )
      + g_{\omega}\omega
      + \frac{g_{\rm A}}{2f_{\pi}}\gamma_{0}\gamma_{5}{\vec\gamma}\cdot
        \tau^{0}\nabla\pi){\psi} = \varepsilon{\psi},
\end{equation}
and the Klein-Gordon equations for the mesons:
\begin{eqnarray}
( -\nabla^2 + {m_{\pi}}^2 )\pi & = & \frac{g_{\rm A}}{2f_{\pi}}\rho_{\rm pv}, \\
( -\nabla^2 + {m_{\sigma}}^2 )\sigma & = &
             - g_{\sigma}\rho_{\rm s}
             - 3\lambda{f_{\pi}}\sigma^2
             - \lambda\sigma^3
             + {\widetilde{g_{\omega}}}^2{f_{\pi}}\omega^2
             + {\widetilde{g_{\omega}}}^2\sigma\omega^2, \\
( -\nabla^2 + {m_{\omega}}^2 )\omega & = &
              g_{\omega}\rho_{\rm v}
            - 2{\widetilde{g_{\omega}}}^2{f_{\pi}}\sigma\omega
            - {\widetilde{g_{\omega}}}^2\sigma^2\omega,
\end{eqnarray}
where we consider the isospin symmetry nucleus, N = Z.  There is a
symmetry theorem for the Hartree-Fock (mean field) approximation
with respect to the symmetry of the original Lagrangian
\cite{ripka,horiuchi}.  In the isospin symmetric nuclear case, we
can verify that the mean field Lagrangian is symmetric under the
isospin rotation to mix the proton and the neutron states. Hence,
we can take a special case, where only $\pi^{0}$ is finite due to
the isospin symmetry of the mean field Lagrangian and write it as
$\pi$. In fact, we have checked this symmetry by performing the
mean field calculations with $\pi^{0}$ in one case and with
$\pi^{\pm}$ in another case and obtained the same energy in both
cases \cite{toki}.  We take the static approximation and assume
the time reversal symmetry of the system. We have introduced here
$g_A$ in the pion nucleon coupling in order to fulfill the
Goldberger-Treiman relation.  In the linear sigma model, we get
$g_A=1$.  In the mean field approximation, the source terms of the
Klein-Gordon equations are replaced by their expectation values in
the ground state.

\begin{eqnarray}
\frac{g_{\rm A}}{2f_{\pi}} \nabla\cdot\bar{{\psi}}\gamma_{5}{\vec\gamma}\tau^{0}{\psi}
& \longrightarrow &
\frac{g_{\rm A}}{2f_{\pi}} \langle
\nabla\cdot\bar{{\psi}}\gamma_{5}{\vec\gamma}\tau^{0}{\psi} \rangle
= \frac{g_{\rm A}}{2f_{\pi}}\rho_{\rm pv}, \\
g_{\sigma}\bar{{\psi}}{\psi}
& \longrightarrow &
g_{\sigma} \langle \bar{{\psi}}{\psi} \rangle  = g_{\sigma}\rho_{\rm s}, \\
g_{\omega}\bar{{\psi}}\gamma_{0}{\psi}
& \longrightarrow &
g_{\omega}\langle \bar{{\psi}}\gamma_{0}{\psi} \rangle = g_{\omega}\rho_{\rm v}.
\end{eqnarray}
The total energy is given by
\begin{eqnarray}
E_{\rm total} & = & \int d^{3}r {\cal H}   \\ \nonumber
              & = & \sum_{njm}\varepsilon_{njm}
                -   \int d^{3}r
     \Big\{  \frac{1}{2}g_{\sigma}\rho_{\rm s}\sigma
           + \frac{1}{2}g_{\omega}\rho_{\rm v}\omega
           - \frac{1}{2}\frac{g_{\rm A}}{2f_{\pi}}\rho_{\rm pv}\pi \\ \nonumber
&&\hspace{4.5cm}
           + \frac{1}{6}(3\lambda{f_{\pi}})\sigma^3
           + \frac{\lambda}{4}\sigma^4
           - \frac{1}{2}{\widetilde{g_{\omega}}}^2f_{\pi}\sigma\omega^2
           - \frac{1}{2}{\widetilde{g_{\omega}}}^2\sigma^{2}\omega^2 \Big\} \\ \nonumber
&&\hspace{7.5cm} - ZM_{\rm p} -ZM_{\rm n} -E_{\rm c.m.},
\end{eqnarray}
where we take the center of mass correction as $E_{\rm c.m.} =
\frac{3}{4}(41A^{\frac{1}{3}})$ MeV. We write here the wave
functions and the densities for the case of the finite pion mean
field.  In this case, the parity of the nucleon is broken, because
the pion source term has the negative parity. The nucleon wave
functions are then written as,
\begin{equation}
\psi_{njmm_{\tau}} = \left(\begin{array}{@{\,}r@{\,}}
   iG_{n\kappa m_{\tau}}{\cal Y}_{\kappa m} \zeta(m_{\tau})
+ iG_{n \bar{\kappa} m_{\tau}}{\cal Y}_{\bar{\kappa} m} \zeta(m_{\tau}) \\
    F_{n\kappa m_{\tau}}{\cal Y}_{\bar{\kappa} m} \zeta(m_{\tau})
+  F_{n \bar{\kappa} m_{\tau}}{\cal Y}_{\kappa m} \zeta(m_{\tau})  \\
\end{array}\right),
\end{equation}
where the summation over $\kappa$ means the parity mixing, where
$\kappa$ is $\kappa=-(l_\uparrow+1)$ for $l_\uparrow=j-1/2$ and
$\kappa=l_\downarrow$ for $l_\downarrow=j+1/2$. Using these wave
functions, we can calculate all the necessary densities as,
\begin{equation}
\rho_{s} = \sum_{nj}W_{nj} \frac{2j+1}{4\pi} \sum_{m_\tau} (
|G_{n\kappa m_\tau }|^2 - |F_{n\kappa m_\tau }|^2
 +|G_{n\bar{\kappa} m_\tau }|^2 -|F_{n\bar{\kappa} m_\tau }|^2 ),
\end{equation}
\begin{equation}
\rho_{v} = \sum_{nj}W_{nj} \frac{2j+1}{4\pi} \sum_{m_\tau} (
|G_{n\kappa m_\tau }|^2 + |F_{n\kappa m_\tau }|^2
 +|G_{n\bar{\kappa} m_\tau }|^2 + |F_{n\bar{\kappa} m_\tau }|^2 ),
\end{equation}
\begin{eqnarray}
\rho_{pv} & = & -2 \sum_{nj}W_{nj} \frac{2j+1}{4\pi} \\
\nonumber & \times & \sum_{m_\tau} (-1)^{\frac{1}{2} - m_{\tau}}
{\Bigl\{ }
  \frac{d}{dr}({G_{n\kappa m_\tau }}^{\ast}G_{n\bar{\kappa} m_\tau })
+ \frac{2}{r}({G_{n\kappa m_\tau }}^{\ast}G_{n\bar{\kappa} m_\tau }) \\
\nonumber && \hspace{5.7cm}+
\frac{d}{dr}({F_{n\kappa m_\tau}}^{\ast}F_{n\bar{\kappa} m_\tau })
+ \frac{2}{r}({F_{n\kappa m_\tau}}^{\ast}F_{n\bar{\kappa} m_\tau })
{\Bigr\} }. \\ \nonumber
\end{eqnarray}
We are now able to calculate the coupled differential equations by
doing iterative calculations.

\begin{figure}[h]
\epsfxsize = 10 cm \centerline{\epsfbox{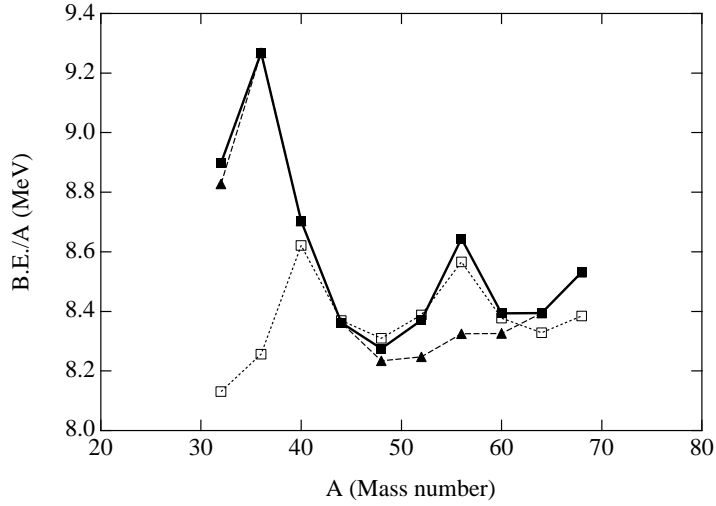}}
\caption[Fig.4]{The binding energy per particle for N = Z
even-even mass nuclei in the neutron number range of N = 16 $\sim$
34. The binding energies per particle for the case of the extended
chiral sigma model without and with the pion mean field are shown
by the dashed and the solid lines.  As a comparison, those for the
RMF(TM1) are shown by the dotted line.}
\end{figure}

In this chapter, we discuss first the properties of finite nuclei
in terms of the extended chiral sigma model without introducing
yet the pion mean field. We show the results of binding energies
per particle of N = Z even-even mass nuclei from N = 16 up to N =
34 in Fig. 4. We take all the parameters of the extended chiral
sigma model as those of the nuclear matter (Figs. 2 and 3) except
for $g_\omega$ = 7.176 instead of 7.033 for overall agreement with
the RMF(TM1) results. For comparison, we calculate these nuclei
within the RMF approximation without pairing nor deformation.  The
RMF(TM1) provides the magic numbers, which are seen as the binding
energy per particle increases at N = Z = 20 and 28. On the other
hand, the extended chiral sigma model without the pion mean field
provides the magic number behavior only at N = Z = 18 instead of N
= Z = 20.

\begin{figure}[h]
\epsfxsize = 10 cm \centerline{\epsfbox{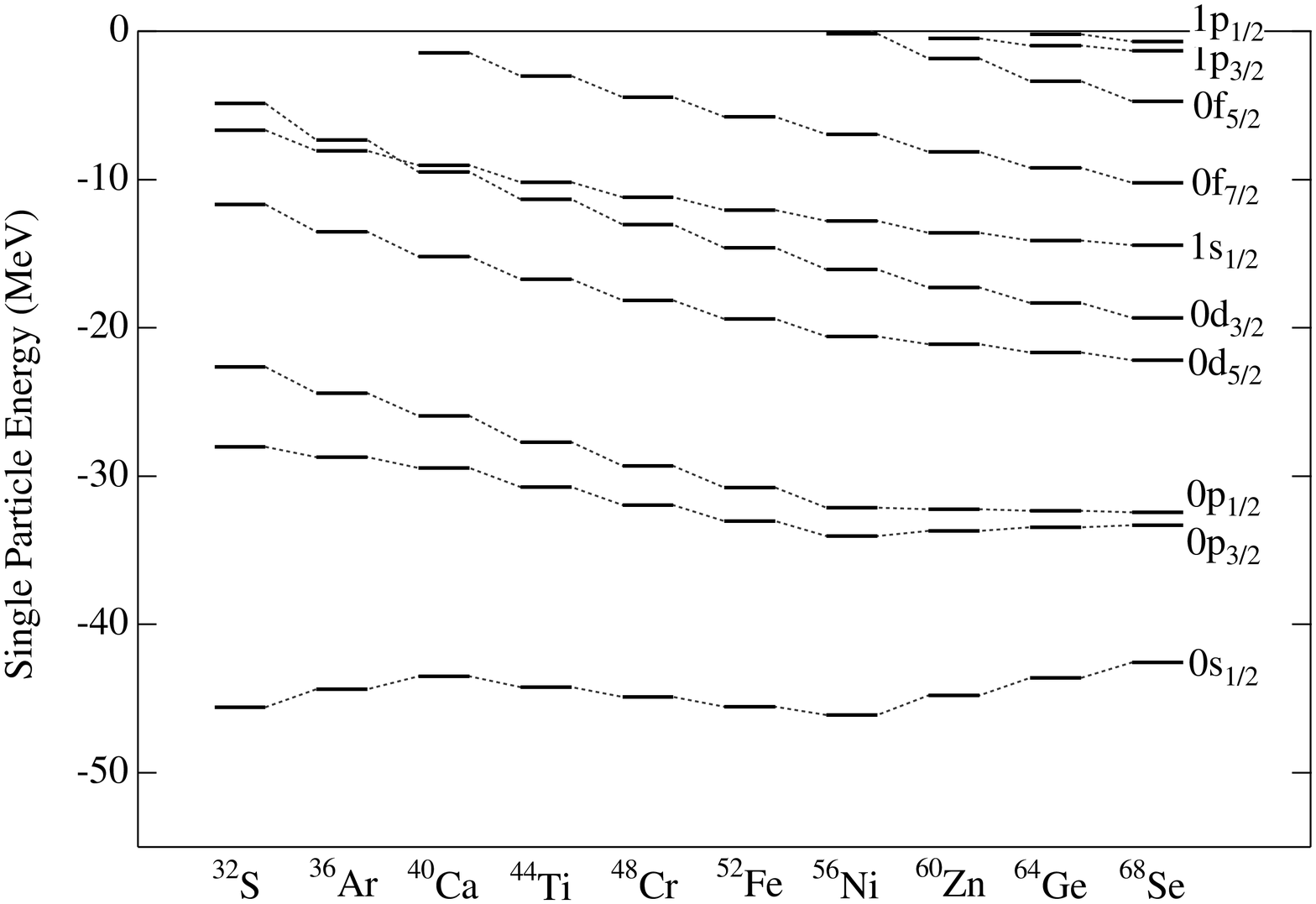}}
\caption[Fig.5]{The proton single particle energies for the N = Z
even-even mass nuclei in the case of the RMF(TM1) theory, where
the magic numbers at N = 20 and 28 are visible. }
\end{figure}
\begin{figure}[h]
\epsfxsize = 10 cm \centerline{\epsfbox{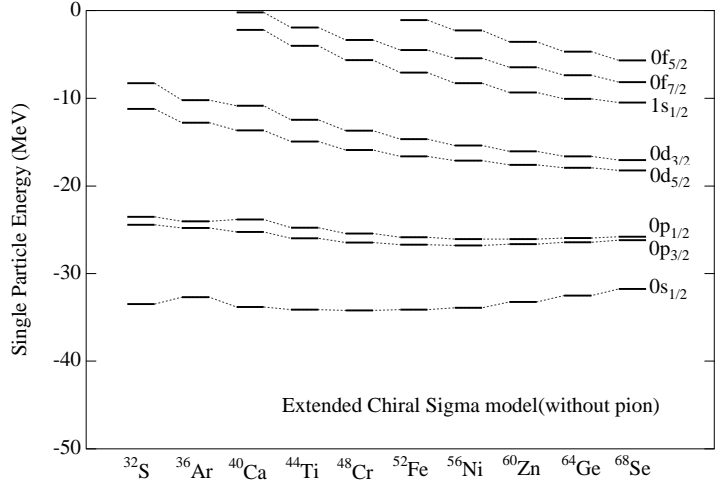}}
\caption[Fig.6]{The proton single particle energies for the N = Z
even-even mass nuclei in the case of the extended chiral sigma
model without the pion mean field. $1s_{1/2}$ orbit is pushed up
and the N = 20 magic number is shifted to N = 18. The spin orbit
splitting between $0f_{7/2}$ and $0f_{5/2}$ is small and the magic
number at N = 28 is not visible.}
\end{figure}

In order to see why the difference between the two models for the
Lagrangian arises, we show in Fig. 5 the single particle levels
for the two models.  In the case of the TM1 parameter set shown in
Fig. 5, the shell gaps are clearly visible at N = 20 and 28.  The
magic number at N = 20 is due to the central potential, while the
magic numbers at N = 28 comes from the spin-orbit splitting of the
0f-orbit. This is definitely due to the fact that the vector
potential and the scalar potential in nuclear matter are large so
as to provide the large spin-orbit splitting.  On the other hand,
the single particle spectrum of the extended chiral sigma model is
quite different from this case as seen in Fig. 6.  Most remarkable
structure is that the $1s_{1/2}$ orbit is strongly pushed up.  Due
to this reason the $0d_{3/2}$ orbit becomes the magic shell at N =
18 and the magic number appears at N = 18 instead of N = 20.  We
see also not strong spin-orbit splitting and hence there appears
no shell gap at N = 28.  The first discrepancy could be due to the
large incompressibility as seen in the nuclear matter energy per
particle as seen in Fig. 2.  The other is due to the relatively
small vector and scalar potentials in nuclear matter as seen in
Fig. 3.

\begin{figure}[h]
\epsfxsize = 10 cm \centerline{\epsfbox{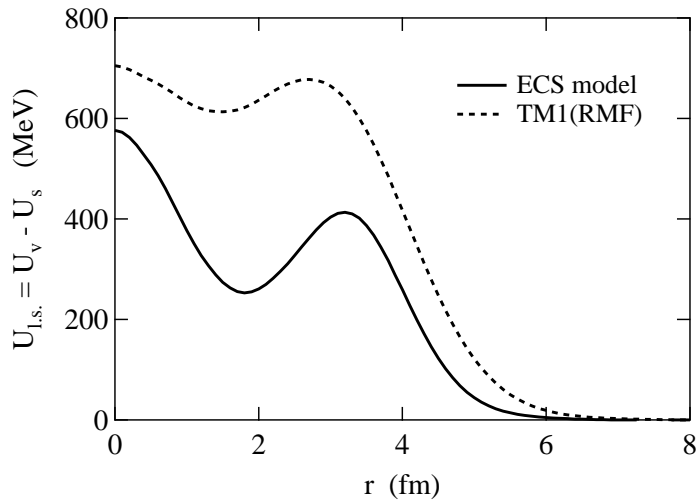}}
\caption[Fig.7]{The scalar-vector potential difference, $U_{ls}$,
which is related with the spin-orbit potential, as a function of
the radial coordinate, r.  The potential difference for the case
of the TM1 parameter set is shown by the dashed curve and for the
ECS model is shown by the solid curve.}
\end{figure}

We would like to detail further the discussion on the spin-orbit
splitting in the situation where the compressibility is very large
as the ECS model, since the spin-orbit splitting is related with
the behavior of the scalar-vector potential difference in the
surface region. We show in Fig. 7 how the scalar-vector potential
difference behaves as a function of r, which is defined as
$U_{ls}=U_V-U_S$. The magnitude of the ECS model is about a half
of the TM1 case. The spin-orbit potential is then defined by
eliminating the small component in the relativistic wave function
and by getting the spin-orbit operator explicitly as,
\begin{equation}
V_{ls}=\frac{2}{r}\frac{\frac{d}{dr}U_{ls}}{(M+\varepsilon-U_{ls})^2}
\vec{l}\cdot\vec{s}
\end{equation}
The spin-orbit potential is proportional to the derivative of the
scalar-vector potential difference, which emphasizes the
contribution from the nuclear surface.  Hence, to compare the
magnitude of the spin-orbit effects of the two cases, we calculate
the volume integrals of $V_{ls}$;
\begin{equation}
\int\frac{2}{r}\frac{\frac{d}{dr}U_{ls}}{(M+\varepsilon-U_{ls})^2}
r^2 dr
\end{equation}
We use for the $\varepsilon$ the value corresponding to the
binding energy of 8MeV, and obtain the ratio of the two cases as
0.48, which is again about a half. Hence, the spin-orbit effect
for the ECS model is about a half of the TM1 case, which could be
seen already in the single particle spectra shown in Fig. 6.

\section{Finite pion mean field for finite nuclei}

We include now the pion mean field in the relativistic mean field
calculation\cite{toki79,oset82}.  The method of the numerical
calculation is provided in the paper of Toki et al.\cite{toki} and
Sugimoto et al.\cite{sugimoto03}. The results on the binding
energy per particle are shown in Fig. 4.  In this calculation we
take 1.15 instead of the experimental axial coupling constant
$g_{\rm A}$ = 1.25 due to the Goldberger-Treimann relation.  We
take this smaller value in order to reproduce the binding energy
for $^{56}$Ni.  It is very interesting to see that the magic
number effect at N = 28 appears as the binding energy per particle
increases at N = 28. This large effect of the finite pion mean
field for the jj-closed shell nuclei has been demonstrated in the
previous work\cite{toki}.

\begin{figure}[h]
\epsfxsize = 10 cm \centerline{\epsfbox{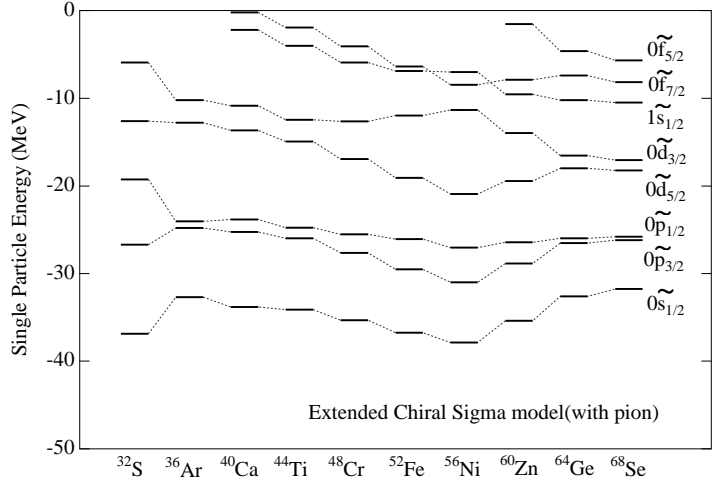}}
\caption[Fig.9]{The proton single particle energies for the N = Z
even-even mass nuclei in the case of the extended chiral sigma
model with the pion mean field.  The spin-orbit splitting is made
large due to the finite pion mean field, which is visible as
centered at the N = Z = 28 nucleus.  We note that while the total
angular momentum is a good quantum number, but the angular
momentum is not exact, we write the dominant angular momentum
beside each single particle state.}
\end{figure}

\begin{figure}[h]
\epsfxsize = 7 cm \centerline{\epsfbox{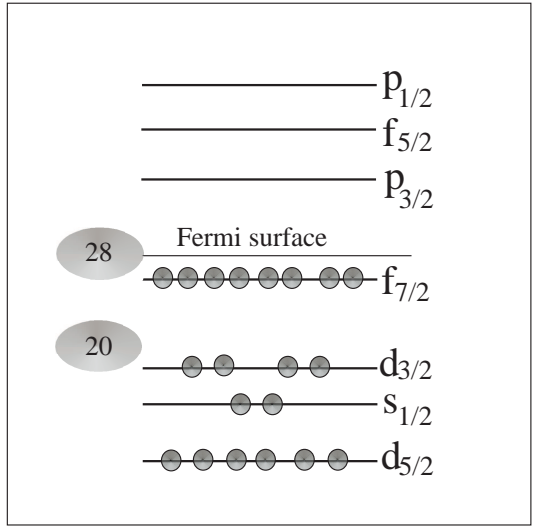}}
\caption[Fig.10]{The schematic picture of the single particle
states and the occupied particles in the $^{56}$Ni nucleus.}
\end{figure}

We give here an intuitive explanation to understand the energy
curve of the magic structure in Fig. 4 to be given by the finite
pion mean field using the schematic picture in Fig. 9. To proceed,
we have to know first the effect of the finite pion mean field in
terms of the shell model.  The discussion of the parity
projection, done in the previous publication\cite{toki}, clearly
shows that the pionic correlations due to the finite pion mean
field is expressed by the coherent 0$^{-}$ particle-hole
excitations, in which the coupling of the different parity states
$l$ and $l'=l \pm 1$ with the same total spin $j$ in the shell
model language.  In the discussion of the contribution to the
pionic correlations from various single particle states, the
highest spin state in each major shell has a special role.  Only
this highest spin state does not find the partner to form the
0$^-$ state in the lower major shells.  However, if this state is
filled by nucleons, those nucleons are able to find the 0$^-$
partners in the higher major shells by making particle-hole
excitations. Hence, the position of the highest spin state in a
major shell with respect to the Fermi surface is important for the
strength of pionic correlations in nuclei.

In the case of discussion, the highest spin state is the $f_{7/2}$
state as shown in Fig. 9. In the $^{40}$Ca case, the occupied
states can not couple with the $f_{7/2}$ state to form 0$^{-}$ and
the $f_{7/2}$ level is not used at all for the pionic
correlations. In the next $^{44}$Ti case, nucleons start to occupy
the $f_{7/2}$ level, and these nucleons are used for the 0$^{-}$
particle-hole excitations into $g_{7/2}$ levels. The number of
particles to be used by the pionic correlation increases as the
nucleon number is increased until $^{56}$Ni, where the $f_{7/2}$
level is completely occupied. For the nuclei above $^{56}$Ni, the
upper shells as $f_{5/2}$ are to be occupied and those states are
not used for 0$^{-}$ particle-hole excitations from the $d_{5/2}$
level below caused by the pionic correlation due to the Pauli
blocking. For $^{56}$Ni, the pionic correlation becomes maximum.
This is the reason why $^{56}$Ni obtains the largest pionic
correlation energy, which leads to the appearance of the magic
number at N = 28.

We discuss now the effect of the finite pion mean field on the
single particle energies. We show in Fig. 8 the single particle
spectra for various nuclei.  We see clearly the large energy
differences between the spin-orbit partners to be produced by the
finite pion mean field as the energy differences become maximum
for nuclei at N = 28. The pion mean field makes coupling of
different parity states with the same total spin.  The $0s_{1/2}$
state repels each other with the $0p_{1/2}$ state and therefore
the $0s_{1/2}$ state is pushed down and the $0p_{1/2}$ state is
pushed up.  The next partner is $0p_{3/2}$ and $0d_{3/2}$.  The
$0p_{3/2}$ state is pushed down, while the $0d_{3/2}$ state is
pushed up.  The next partner is $0d_{5/2}$ and $0f_{5/2}$.  The
$0d_{5/2}$ state is pushed down, while the $0f_{5/2}$ state is
pushed up.  This pion mean field effect continues to higher spin
partners.  This coupling of the different parity states with the
same total spin due to the finite pion mean field causes the
splittings of the spin-orbit partners as seen clearly for the $0p$
spin-orbit partner, $0d$ spin-orbit partner and $0f$ spin-orbit
partner in $^{56}$Ni.  It is extremely interesting to see that the
appearance of the energy splitting between the spin-orbit partners
for the case of the finite pion mean field is caused by completely
a different mechanism from the case of the spin-orbit interaction.

We would like to see the contributions of each term in the
Lagrangian for the cases with and without the pion mean field in
Table 1.
\begin{table}
\caption{The binding energy per particle (BE/A) and the
contributions of the sum of sigma and omega, (U$_{\sigma} +$
U$_\omega$), kinetic (KE), pion (U$_\pi$), non-linear term (NL),
sigma-omega coupling term (CP) and Coulomb (U$_{\rm C}$) energies
per nucleon in MeV for $^{56}$Ni in the extended chiral sigma
model.}
\vspace{5mm}
\begin{tabular}{cccccccc}
\hline \hline
  & BE/A   &   ${\rm U}_{\sigma} + {\rm U}_{\omega}$   & KE
  & ${\rm U}_{\pi}$   &   NL  &  CP  & ${\rm U}_{\rm C}$ \\ \hline
with $\pi$ field      & 8.6  & -21.8  & 20.9  & -2.9  & 8.1  &
-15.4 & 2.6 \\ \hline without $\pi$ field   & 8.4  & -22.6  & 18.8
&     0 & 8.0  & -15.0 & 2.6 \\ \hline \hline
\end{tabular}
\end{table}

The binding energy increases slightly by making the pion mean
field finite.  The pion term contributes attractively and the
energy gain due to the pion term is obtained by making the kinetic
energy and the sum of the sigma and omega potential terms
increase.  The structure of the wave functions changes largely,
while the total energy is kept almost unchanged.  This change of
the structure will make the observables associated with the spin
quantities change largely.  The effect of the structure change on
various observables will be studied in the near future.

\section{Conclusion}\label{sec: conc}

We have studied infinite nuclear matter and finite nuclei with the
nucleon number N = Z even-even mass in the range of N = 16
and N = 34 using the chiral sigma model, which is good
for hadron physics. The direct application of the chiral sigma
model is not able to provide the good saturation property of
infinite matter. We have then used the extended chiral sigma (ECS)
model, in which the omega meson mass is dynamically generated by
the sigma condensation as the nucleon mass.
This ECS model is able to provide a good saturation property, although the
incompressibility comes out to be too large.
Another characteristic property of the ECS model is that the scalar
and vector potentials are about a half of the case of the RMF(TM1)
model in nuclear matter.

We have then applied this ECS model to finite nuclei. The ECS
model without the pion mean field gives the result that the magic
number appears at N = 18 not at N = 20.  This result comes from
the large incompressibility found in the equation of state as K =
650 MeV.  This property of the ECS model provides the mean field
central potential repulsive in the interior region and the
1$s$-orbit is extremely pushed up. Due to this, the magic number
appears at N = 18 instead of N = 20. We note that this problem
originates from the ECS model treated in the present framework and
the finite pion mean field under the mean field approximation does
not remove this difficulty. There are several possibilities to be
worked out to cure this problem as the effect of Dirac sea, the
parity projection, and the Fock term.

The ECS model without the pion mean field provides the result
that the magic number does not appear at N = 28. This result comes
from another characteristic property of the ECS model, which is
the small scalar and vector potentials in nuclear matter. The
scalar and vector potentials lead directly to the strength of the
spin-orbit interaction in finite system. Since the spin-orbit
interaction given by the ECS model is about a half of those of the
standard RMF calculation with the TM1 parameter set, the energy
splittings between the spin-orbit partners are small and,
therefore, there appears no magic effect at N = 28. As for this
point, it is important to introduce the pion mean field by
breaking the parity of the single particle states in the ECS model
Lagrangian. Since the role of the pion mean field on the jj-closed
shell nuclei has been demonstrated in the previous
publication\cite{toki}, we have introduced the parity mixed
intrinsic single particle states in order to treat the pion mean
field in finite nuclei.  We followed the formulation of Sugimoto
et al.\cite{sugimoto03} in the RMF framework. We have found that
the magic number effect appears at N = 28.  We have studied the
change of the single particle spectrum due to the finite pion mean
field. It is extremely interesting to find that the spin-orbit
partners are split largely by the pion mean field effect.  Namely,
the parity partners as ($s_{1/2}$ and $p_{1/2}$), ($p_{3/2}$ and
$d_{3/2}$) and ($d_{5/2}$ and $f_{5/2}$) are pushed out each other
due to the pion mean field and as the consequence the spin-orbit
partners are split largely like the ones of the ordinary
spin-orbit splittings. This is related with the energy differences
of the spin-orbit partners caused by the energy loss of the tensor
(pionic) correlations due to the Pauli blocking\cite{terasawa60}.

It is gratifying to observe that first the extended chiral sigma
model, which has the chiral symmetry and its dynamical symmetry
breaking, is able to provide the nuclear property with only a
small adjustment of the parameters in the Lagrangian. The energy
splitting between the spin-orbit partners appears remarkably in
the ECS model with the pion mean field. The most important
consequence obtained in this study is that this energy splitting
is caused by the pion mean field which is completely a different
mechanism from the case of the spin-orbit interaction introduced
phenomenologically. This suggests the origin of the magic effect
of jj-closed shell nuclei.

\section*{Acknowledgement}
We acknowledge fruitful discussions with Prof. Y. Akaishi, Prof.
H. Horiuchi and Prof. I. Tanihata on the roll of the pion in
nuclear physics. We are grateful to Dr. D. Jido for helpful
discussions on the linear sigma model.  We thank Prof. E. Oset for
reading manuscript and valuable discussions. This work is
supported in part by the Grant-in-aid for Scientific Research
(B) 14340076 of the Ministry of Education, Culture, Sports, Science
and Technology of Japan.


\end{document}